# Yield stress and other flow and wall slip parameters of viscoplastic fluids from steady torsional flow*

Dilhan M. Kalyon**

Stevens Institute of Technology,

1 Castle Point St. Hoboken, NJ 07030, USA

**Abstract**

The ubiquitous wall slip behavior of viscoplastic fluids renders the analysis of their steady torsional flow data to determine their yield stress and other parameters of their shear viscosity material function challenging. Roughened surfaces to prevent wall slip can be used but such methods frequently result in the fracturing of the sample. Instead, it is shown here that the yield stress of a viscoplastic fluid can be determined from the torque versus apparent shear rate data of steady torsional flow using a single gap. The step change in the slope of the torque versus apparent shear rate data of a viscoplastic fluid marks the yield stress of the fluid. For demonstration, the earlier-characterized parameters of the shear viscosity and apparent wall slip behavior of a viscoplastic suspension [He *et al.*, J. Rheology, 63, 19 (2019)] were used for obtaining the shear stress distribution and the torque at each rotational speed. The yield stress, determined from the torque versus apparent shear rate data, could be verified by using a second method relying on the comparison of the wall slip velocities with wall velocities, while at the same time enabling the determination of the other parameters of the shear viscosity material function of the viscoplastic fluid.

I. **Introduction**

The steady torsional flow using parallel plates (disks) is widely employed for the characterization of the shear viscosity and wall slip velocity versus shear stress behavior of complex fluids. One common denominator for the flow and deformation behavior of various complex fluids, including concentrated suspensions of rigid particles or microgels with soft particles, is their viscoplastic nature in flow. Viscoplastic fluid models including Bingham, Casson, Herschel-Bulkley and Robertson-Stiff are generalized Newtonian fluids which are based on the stipulation that under steady flow conditions the rate of deformation would be zero when the stress magnitude is less than the yield stress [1-3]. Such classical viscoplastic fluid models were used in conjunction with the no-slip condition at the wall. However, similar to the behavior of many other complex fluids, viscoplastic fluids exhibit wall slip [4-16]. A logical extension of the viscoplastic constitutive equations has been to apply them also to cases where there is slip at the wall [17-24].

For viscoplastic fluids the concomitant characterizations of the wall slip behavior and the parameters of the shear viscosity material function, including the determination of the yield stress value, are important challenges. Flow visualization methods, including magnetic resonance imaging, MRI, [25-32], microscopy in conjunction with a transparent window and a high speed camera [33], particle tracking velocimetry, PTV [34], and particle image velocimetry, PIV, [34-38] were utilized for obtaining the velocity distributions and the parameters of the viscoplastic constitutive equations in various steady viscometric flows. For example it has been demonstrated that PIV, when used in conjunction with capillary rheometry, Couette flow or axial annular flow, enables the characterization of the velocity distributions and provides the flow curves under fully-developed, and isothermal conditions to unambiguously generate the parameters of wall slip behavior and the shear viscosity of viscoplastic fluids [35-38]. However, the applications of rheometers with flow visualization capabilities are handicapped by the constraints in the types of fluids that can be characterized, the requisite nature of the rheometer walls and the resolutions of such methods for the determination of the velocities at the walls. For example, MRI requires non-metallic rheometer walls and PIV can only be used for the characterization of transparent fluids in viscometers with transparent walls.

Without the benefit of such flow visualization methods the characterization of the shear viscosity material function of various complex fluids, including viscoplastic fluids, can be accomplished based on the characterization of the wall slip velocity versus the wall shear stress [39]. The method involves the determination of the wall slip velocity values as a function of shear stress, followed by the correction of the flow curves (wall shear stress versus the apparent shear rate) to generate the true "wall slip-corrected" shear rates, as well as the yield stresses and other parameters of the shear viscosity of viscoplastic fluids [5, 6, 20]. Such data analysis and correction methods were applied to data obtained using multiple rheometers [6, 36, 41-42] and the characterized shear viscosity and wall slip behavior were found to be consistent with different rheometers. Wall slip behavior can be used as the flow boundary condition in the mathematical modeling and simulation of the thermo-mechanical histories of viscoplastic fluids in complex flows [12, 17, 18, 37, 45]. For example, the analysis of extrusion flows in conjunction with the wall slip boundary conditions, determined from viscometric flows, can be carried out to obtain the pressure and temperature distributions in continuous processors [12, 17, 18, 37, 45].

However, these procedures are not widely used for the concomitant characterization of the wall slip behavior and shear viscosity material function of viscoplastic fluids, possibly due to the generally labor-intensive nature of the data collection and analysis methods. Generally, roughened rheometer surfaces (or surfaces with protrusions) are used in empirical attempts to eliminate wall slip, frequently leading to the fracture or shear banding of viscoplastic samples, while the opportunity to characterize the wall slip behavior of the viscoplastic fluid is wasted [7, 8, 19]. Especially, for processability analysis it is often overlooked that if a viscoplastic fluid exhibits significant slip during rheological characterization, it will also exhibit wall slip during its processing flow [17]. A good understanding of the wall slip behavior from viscometric flows can help solve important processing problems [12, 17, 18, 40, 41, 45]. Here, the determination of the yield stress, based on the analysis of the torque and apparent shear rate data from steady torsional flow using smooth surfaces, is demonstrated. The yield stress determination is followed by the characterization of the other parameters of the shear viscosity material function as well as the wall slip behavior of the viscoplastic fluid.

## II. Background: Steady torsional flow

The schematics of steady torsional flow involving two parallel and smooth plates (disks) with radius, R, the top one of which is rotating at a rotational speed of Ω rad/s and the other is stationary, with the fluid sandwiched in between the two disks, is shown in Fig. 1. During the experiments which are carried out under steady, isothermal and laminar flow conditions (typically creeping flow for high-molecular weight polymer melts, concentrated suspensions and microgels), either a torque is applied and the corresponding rotational speed Ω is determined, or a rotational speed of Ω is imposed and the corresponding steady torque is determined.

Both methods can be used in conjunction with systematic changes in the surface to volume ratio of the specimen held in between two disks to characterize wall slip velocities [5-7]. The wall slip velocities can then be used to obtain true flow curves (true shear stress versus true shear rate) for the characterization of the shear viscosity of the viscoplastic fluid. The wall slip velocity analysis can be also used to enable the determination of the yield stress of viscoplastic fluids [5, 17, 20, 36-38]. In this method the shear stress at which the transition from plug flow to flow which sustains a continuous deformation rate becomes evident from the inspection of the wall slip velocity at the wall versus the shear stress (and comparison with the wall velocity for steady torsional flow, or the mean velocity in channel flows) to provide the yield stress value of the viscoplastic fluid [36-38]. However, data need to be collected at multiple gaps to be able to carry out the Mooney analysis for determination of wall slip velocities [39]. In the following an additional method for the determination of the yield stress, involving only the inspection of the torque versus the apparent shear rate data is demonstrated for the steady torsional flow of a concentrated suspension with apparent wall slip.

**Analysis of the steady torsional flow (parallel-disk rheometer)**
Let us consider the steady state, isothermal, inertia-free creeping flow of an incompressible, viscoplastic fluid sandwiched in between two disks, the upper one at $z=H$ is rotating, and the bottom one at $z=0$ is stationary. The flow is free of particle and binder migration and fracture effects. For relatively small gaps, $H$, i.e., $H<<R$, the shear stress $\tau_{z\theta}(r)$ can be assumed to be constant across the gap at any radial position, $r$ and $\tau_{z\theta} >> \tau_{r\theta}$ [2]. Wall slip velocities, defined as the difference between the velocity of the fluid at the wall and the velocity of the wall, prevail at both walls, i.e., the viscoplastic fluid exhibits wall slip velocities, $U_s(r,0)$ and $U_s(r,H)$ at the surfaces of the disks at the bottom and the top, i.e., z=0 and z=H, respectively. If the two walls are constructed out of similar materials of construction and roughness, and if there are no gravity-driven effects like sedimentation or particle migration, the absolute values of the wall slip velocities would be similar at the two walls, since both walls experience the same shear stress. The shear stress, $\tau_{z\theta}(r)$, is negative for the considered case of the upper wall moving (with the sign convention of stresses assumed to be positive in compression) (Fig. 1).

These considerations result in the following equality for any radial position, $r$:

$$\frac{\Omega r}{H} = \frac{U_s(r,0)}{H} - \frac{U_s(r,H)}{H} + \frac{dV_\theta}{dz}(r) \tag{1a}$$

and at the edge, $r = R$:
$$\frac{\Omega R}{H} = \frac{U_s(R,0)}{H} - \frac{U_s(R,H)}{H} + \frac{dV_\theta}{dz}(R) \tag{1b}$$

where $\frac{\Omega r}{H}$ is the apparent shear rate, $\dot{\gamma}_{ar}$, at the radial position, r, $\frac{\Omega R}{H}$ is the apparent shear rate at the edge, $\dot{\gamma}_{aR}$, and $\frac{dV_\theta}{dz}(R)$ is the true shear rate, $\dot{\gamma}(R)$, imposed on the fluid at $r=R$, i.e., corresponding to the shear stress at the edge, $\tau_{z\theta}(R)$. The slip velocity at the wall, i.e., the difference between the velocity of the fluid and the wall velocity, is negative for the top disk which is moving, i.e., $U_s(r,H)<0$, and the wall slip velocity, $U_s(r,0)$, is positive for the bottom disk, which is stationary, i.e., $U_s(r,0)>0$. The wall slip velocities at the top and bottom disks are related to each other as $U_s(r,H) = -U_s(r,0)$. To avoid confusion "slip velocity" will refer to the absolute values of the wall slip velocities at the top and bottom surfaces.

The dependence of the shear stress, $|\tau_{z\theta}(r)| = -\tau_{z\theta}(r)$, on radial position indicates that for viscoplastic fluids part of the flow field in between the two disks, for which $|\tau_{z\theta}(r)| \leq \tau_0$, will involve plug flow, i.e., $\dfrac{dV_\theta}{dz}(r \le r_0) = 0$, and part of the flow field which experiences $|\tau_{z\theta}(r)| > \tau_0$ will involve finite deformation rates, i.e., $\dfrac{dV_\theta}{dz}(r > r_0) \ne 0$. The radial position $r=r_0$ designates the location at which $|\tau_{z\theta}(r_0)| = \tau_0$. The entire flow field would be plug flow if $|\tau_{z\theta}(R)| \le \tau_0$ (i.e., $r_0 > R$).

The torque, $\mathfrak{I}$, that is necessary to rotate the upper disk at a given apparent shear rate at the edge, $\dot\gamma_{aR}$, is given by:

$$\mathfrak{I} = 2\pi \int_0^R \left(-\tau_{z\theta}(r)\right) r^2 dr \qquad (2a)$$

Upon a change of variable of integration to:

$$\dfrac{\dot\gamma_{aR}^{\,3}\,\mathfrak{I}}{2\pi R^3} = \int_0^{\dot\gamma_{aR}} \left(-\tau_{z\theta}(r)\right)\dot\gamma_{ar}^{\,2}\, d\dot\gamma_{ar} \qquad (2b)$$

and differentiation with respect to the apparent shear rate at the edge, $\dot\gamma_{aR} = \Omega R / H$, and utilizing the Leibniz rule of integration, one obtains [5]:

$$-\tau_{z\theta}(R) = \dfrac{\mathfrak{I}}{2\pi R^3}\left(3 + \dfrac{d\ln \mathfrak{I}}{d\ln(\Omega R / H)}\right) \qquad (3a)$$

For a Newtonian fluid, the derivative $\dfrac{d\ln \mathfrak{I}}{d\ln(\Omega R / H)}$ is equal to one so that the apparent shear stress at the edge becomes:

$$-\tau_{z\theta}(R) = -\tau_{z\theta,app}(R) = \dfrac{2\mathfrak{I}}{\pi R^3} \qquad (3b)$$

Most commercial rheometers use the Newtonian expression for shear stress given in Eq. (3b). The error of using the Newtonian expression (Eq. (3b)) in the determination of the shear stress at the edge is between -15% to 10% for slope $\dfrac{d\ln \mathfrak{I}}{d\ln(\Omega R / H)}$ equal to 0.4 and 1.4, respectively.

Eq. (3a) indicates that the availability of a set of torque data, $\Im$, collected upon the imposition of a set of apparent shear rates at the edge, $\frac{\Omega R}{H}$, would provide the slope $\frac{d\ln \Im}{d\ln(\Omega R/H)}$ so that the shear stress at the edge values $|\tau_{z\theta}(R)| = -\tau_{z\theta}(R)$ can be determined via Eq. (3a). If the data are collected as a function of the surface to volume ratio of the rheometer, i.e., function of the gap between the two disks, the wall slip velocities and the true deformation rates at the edge, $\dot{\gamma}(R)$, can be obtained as a function of $|\tau_{z\theta}(R)|$. Yoshimura and Prud'homme have suggested that the slope $\frac{d\ln \Im}{d\ln(\Omega R/H)}$ would be a function of the gap in between the two disks [5].

Below it will be shown first that the fundamental behavior of the viscoplastic fluid dictates that there be multiple slopes $\frac{d\ln \Im}{d\ln(\Omega R/H)}$ (even for the case of data collected at constant gap, $H$) complicating the conversion of the torque data, $\Im$, to shear stress at the edge, $|\tau_{z\theta}(R)|$. Second, it will be shown that the presence of multiple slopes in torque data (below and above the yield stress value of the viscoplastic fluid) facilitates the determination of the yield stress of a viscoplastic fluid.

### III. Determination of the yield stress values of viscoplastic fluids:

The constitutive equations proposed for viscoplastic fluids generally fit the following [2]. The deformation rate (as represented by the rate of deformation tensor, $\underline{\Delta}$) is zero for conditions under which the $1/2(\underline{\tau}:\underline{\tau}) < \tau_0^2$ i.e.,

$$\underline{\Delta} = 0 \quad \text{for } 1/2(\underline{\tau}:\underline{\tau}) \leq \tau_0^2 \qquad (4a)$$

and for the condition $1/2(\underline{\tau}:\underline{\tau}) > \tau_0^2$ the generalized Newtonian fluid (GNF) is used, i.e.,

$$\underline{\tau} = -\eta(II_\Delta)\underline{\Delta} \quad \text{for } 1/2(\underline{\tau}:\underline{\tau}) > \tau_0^2 \qquad (4b)$$

where the shear viscosity, $\eta$, is a function of the second invariant of the rate of deformation tensor, $II_\Delta$, i.e., $\eta(II_\Delta)$. It was shown for various viscometric and processing flows that the Herschel-Bulkley equation accurately represents the behavior of various viscoplastic fluids for $1/2(\underline{\tau}:\underline{\tau}) > \tau_0^2$ [2, 12, 36- 38, 42-46], i.e.,

$$\underline{\tau} = -\left(\frac{\tau_0}{\left|\sqrt{1/2(\underline{\Delta}:\underline{\Delta})}\right|} + m\left|\sqrt{1/2(\underline{\Delta}:\underline{\Delta})}\right|^{n-1}\right)\underline{\Delta} \quad \text{for } 1/2(\underline{\tau}:\underline{\tau}) > \tau_0^2 \qquad (4c)$$

The Herschel-Bulkley equation involves three parameters at constant temperature, i.e., the yield stress, $\tau_0$, the consistency index, $m$, and the shear rate sensitivity index, $n$. For steady torsional flow the Herschel-Bulkley equation becomes (Fig. 1a and 1b):

$$\tau_{z\theta}(r) = \pm\tau_0 - m\left|\frac{dV_\theta}{dz}(r)\right|^{n-1}\left(\frac{dV_\theta}{dz}(r)\right) \quad \text{for} \quad |\tau_{z\theta}(r)| > \tau_0 \qquad (5a)$$

$$\frac{dV_\theta}{dz}(r) = 0 \quad \text{for} \quad |\tau_{z\theta}(r)| \leq \tau_0 \qquad (5b)$$

In Equation 5a the – sign is used when the shear stress, $\tau_{z\theta}(r)$ is negative. Considering the case of the top disk rotating (Fig. 1), so that $\tau_{z\theta}(r) < 0$, Eq. (5a) becomes:

$$\tau_{z\theta}(r) = -\tau_0 - m\left|\frac{dV_\theta}{dz}(r)\right|^{n-1}\left(\frac{dV_\theta}{dz}(r)\right) = -\tau_0 - m\left(\frac{dV_\theta}{dz}(r)\right)^n \qquad (5c)$$

Fig. 1a and 1b show the schematics of the velocity distributions for viscoplastic fluids under the condition of the shear stress, $|\tau_{z\theta}(r)| \leq \tau_0$ and $|\tau_{z\theta}(r)| > \tau_0$, giving rise to plug flow occurring in between the two walls and a deformation rate imposed on the fluid, respectively. Slip at the wall prevails under both conditions. For $|\tau_{z\theta}(r)| \leq \tau_0$, i.e., for plug flow, the slip velocity (velocity of the fluid at the wall minus the velocity of the wall) is only a function of the plate velocity, $\Omega r$ [20],

$$U_s(r,0) = \frac{\Omega r}{2} \quad \text{and} \quad U_s(r,H) = -\frac{\Omega r}{2} \qquad (6)$$

Equation (6) indicates that under plug flow conditions the wall slip velocities at the bottom, z=0, and at the top, z=H, are independent of the gap, H, between the two disks. The wall slip behavior of various viscoplastic fluids, including concentrated suspensions and gels with rigid and soft particles, is generally subject to the apparent slip mechanism. This mechanism will be used in our demonstration and is reviewed next.

**Apparent wall slip mechanism**

During the flow of a suspension of rigid or soft particles the particles cannot physically occupy the space adjacent to a wall as efficiently as they can away from the wall. This leads to the formation of a generally relatively thin, but always-present, layer of pure fluid adjacent to the wall, i.e., the "apparent slip layer" or the "Vand layer" [47]. For suspensions of rigid particles the estimates of the slip layer thickness over the particle diameter ratio are available [6, 14, 15, 20, 48-50]. Meeker *et al.* have shown that the apparent slip mechanism is also applicable to microgel pastes and concentrated emulsions and have provided methods for the estimation of the apparent slip layer thickness, $\delta$, based on elastohydrodynamic lubrication between squeezed soft particles and shearing surfaces [14, 15]. For viscoplastic microgels the apparent slip mechanism could be integrated into the analysis of various flows including steady torsional, capillary, tangential annular (Couette), axial annular and vane in cup flows [36, 37, 38, 46].

The relationship between the slip velocity, $U_s(\tau_{z\theta}(r))$, and the shear stress, $\tau_{z\theta}(r)$, for apparent wall slip occurring in steady torsional flow becomes the following for Vand layers, the shear viscosity of which can be represented by a power law equation represented with a consistency index, $m_b$, and a power law index of $n_b$, i.e., $\tau_{z\theta}(r) = -m_b \left|\dfrac{dV_\theta}{dz}(r)\right|^{n_b-1}\left(\dfrac{dV_\theta}{dz}(r)\right) = -m_b\left(\dfrac{dV_\theta}{dz}(r)\right)^{n_b}$, as:

$$U_s(r,0) = \dfrac{\delta}{m_b^{1/n_b}}\left(-\tau_{z\theta}(r)\right)^{1/n_b} \text{ and } U_s(r,H) = -\dfrac{\delta}{m_b^{1/n_b}}\left(-\tau_{z\theta}(r)\right)^{1/n_b} \quad (7)$$

Thus, in conjunction with the apparent slip at the wall mechanism the shear stress $\tau_{z\theta}(r)$ at any radial position, *r*, can be determined for the case of the top surface moving using [20]:

$$\left[\frac{-(\tau_{z\theta}(r)+\tau_0)}{m}\right]^{1/n}\left(1-\frac{2\delta}{H}\right)+\frac{2\delta}{H}\left(\frac{-\tau_{z\theta}(r)}{m_b}\right)^{1/n_b}=\frac{\Omega r}{H} \quad \text{for } -\tau_{z\theta}(r) > \tau_0 \quad (8a)$$

$$\frac{2\delta}{H}\left(\frac{-\tau_{z\theta}(r)}{m_b}\right)^{1/n_b}=\frac{\Omega r}{H} \quad \text{for } -\tau_{z\theta}(r) \leq \tau_0 \quad (8b)$$

Equations 8a and 8b indicate that the relationship between the shear stress, $|\tau_{z\theta}(r)|$ and the apparent shear rate expected for the pure plug flow, i.e., for $-\tau_{z\theta}(r) \leq \tau_0$ (Eq. (8b)) would be different than the one that prevails under shear stresses for which $-\tau_{z\theta}(r) > \tau_0$ (Eq. (8a)). How would this manifest itself for the torque, $\Im$ versus the apparent shear rate at the edge, $\Omega R/H$, behavior and how different would be the slope $\frac{d\ln\Im}{d\ln(\Omega R/H)}$ for the continuous deformation rate region, i.e., $|\tau_{z\theta}(r)| > \tau_0$ in comparison to the plug flow region, i.e., $|\tau_{z\theta}(r)| \leq \tau_0$?

For the apparent wall slip mechanism the torque values for pure plug flow, i.e., $|\tau_{z\theta}(R)| \leq \tau_0$, can be determined as a function of the apparent shear rate at the edge of the disks, $\dot{\gamma}_{aR} = \Omega R/H$, as the following for a binder with a power-law type shear viscosity with the consistency index $m_b$ and a power-law index $n_b$:

$$\Im(r_0 > R) = \frac{2\pi m_b R^3}{(3+n_b)}\left(\frac{\Omega R}{2\delta}\right)^{n_b} = \frac{2\pi m_b R^3 H^{n_b}}{(3+n_b)(2\delta)^{n_b}}(\dot{\gamma}_{aR})^{n_b} \quad (9a)$$

Eq. (9a) indicates that the torque for the pure plug flow would be independent of the gap, $H$, regardless of whether the binder fluid is Newtonian or non-Newtonian (note that the product $H\dot{\gamma}_{aR} = \omega R$, thus there is no gap dependence). For a non-Newtonian binder that constitutes the apparent slip layer, with shear viscosity represented by a power-law equation the slope $\dfrac{d \ln \Im}{d \ln (\Omega R / H)}$ would be equal to the power law index of the binder, $n_b$. For a Newtonian binder with viscosity, $\mu_b$.

$$\Im(r_0 > R) = \frac{\pi \mu_b R^4 \Omega}{4\delta} = \frac{\pi \mu_b R^3 H}{4\delta} \dot{\gamma}_{aR} \tag{9b}$$

and hence $\dfrac{d \ln \Im}{d \ln (\Omega R / H)} = 1$.

For cases where there is simultaneous plug flow and continuous deformation occurring simultaneously:

$$\Im = 2\pi \int_0^R (-\tau_{z\theta}(r)) r^2 dr = 2\pi \left[ \int_0^{r_0} (-\tau_{z\theta}(r)) r^2 dr + \int_{r_0}^R (-\tau_{z\theta}(r)) r^2 dr \right] \tag{10a}$$

where $r_0$ is the radial location at which $-\tau_{z\theta}(r_0) = \tau_0$. The first term on the right is thus the contribution of the flow in the plug flow zone to the torque and is equal to:

$$2\pi \int_0^{r_0} (-\tau_{z\theta}(r)) r^2 dr = \frac{2\pi m_b r_0^{3+n_b}}{(3+n_b)} \left( \frac{\Omega}{2\delta} \right)^{n_b} \tag{10b}$$

An analytical expression for the second integral on the right of Eq. (10a) can be substituted if in the continuous deformation region of steady torsional flow i.e., for $-\tau_{z\theta}(r) > \tau_0$, the slip contribution to shear stress is assumed to be negligible in comparison to the contribution of bulk deformation of the suspension to shear stress:

$$\int_{r_0}^R (-\tau_{z\theta}(r)) r^2 dr = \frac{\tau_0}{3} (R^3 - r_0^3) + \frac{m}{(3+n)} \left( \frac{\Omega}{H} \right)^n (R^{3+n} - r_0^{3+n}) \tag{10c}$$

so that the torque, $\Im$, for the condition of negligible slip contribution in the continuous deformation region becomes:

$$\Im = 2\pi \int_0^R (-\tau_{z\theta}(r)) r^2 dr = 2\pi \left[ \frac{m_b r_0^{3+n_b}}{(3+n_b)} \left( \frac{\Omega}{2\delta} \right)^{n_b} + \frac{\tau_0}{3} (R^3 - r_0^3) + \frac{m}{(3+n)} \left( \frac{\Omega}{H} \right)^n (R^{3+n} - r_0^{3+n}) \right]$$
$$\tag{10d}$$

The third term on the right side of Eq. (10d) dominates for $R \gg r_0$ so that $\frac{d\ln \Im}{d\ln(\Omega R/H)} \approx n$ ($n$ being the shear rate sensitivity index of the Herschel-Bulkley fluid, Eq. (5a-c)).

Thus, in the steady torsional flow of viscoplastic fluids subject to apparent wall slip a slope change in $\frac{d\ln \Im}{d\ln(\Omega R/H)}$ from the power law index of the binder, $n_b$ to a value approaching $n$ (on the smaller side considering the role wall slip plays in lowering the torque, $\Im$) is expected for relatively high shear stress at the edge values. The change in slope reflects the transition from pure plug flow to a flow with both plug flow for $r<r_0$ and continuous deformation for $r_0<r<R$. The change is slope, $\frac{d\ln \Im}{d\ln(\Omega R/H)}$, is expected to occur when the shear stress at the edge becomes equal to the yield stress, i.e., $|\tau_{z\theta}(R)| = \tau_0$. Thus, this step change in the slope $\frac{d\ln \Im}{d\ln(\Omega R/H)}$ serves as the basis for the determination of the yield stress, $\tau_0$, values of viscoplastic fluids using steady torsional flow. It is sufficient to collect torque, $\Im$, versus rotational speed, $\Omega$, data at a single gap, H.

In the following this method for the determination of the yield stress is demonstrated on a viscoplastic concentrated suspension characterized by He et al. [42]. In the He et al. investigation a series of concentrated suspensions were prepared using both batch and continuous mixers in the solid volume fraction $\phi$, range of 0.62 to 0.82. The binder was a poly(dimethyl siloxane), PDMS with a molecular weight of 48 kDa. The shear viscosity of the PDMS was independent of the shear stress up to 3,000 Pa, exhibiting a zero-shear-rate viscosity of 4.9 Pa-s, so that $n_b = 1$ and $m_b = \mu_b = 4.9$ Pa-s for $|\tau_{z\theta}(r)| < 3000$ Pa [42]. PDMS itself does not exhibit wall slip for molecular weights < 500 kDa or when the wall shear stress < 70 kPa [51]. He et al. have indeed confirmed that the PDMS binder used does not slip at the wall [42]. The particle radii, $a > 1\mu m$. The typical particle Reynolds number, Re, and the Peclet number, Pe, were determined to be $\text{Re}(\dot{\gamma}) = \frac{\rho a^2 \dot{\gamma}}{\mu_b} \ll 1$, $Pe(\dot{\gamma}) = \frac{6\pi\mu_b a^3 \dot{\gamma}}{kT} \gg 1$, suggesting that the steady torsional flow took place in the creeping flow regime and that the viscous forces dominated over the colloidal forces over the entire shear rate, $\dot{\gamma}$ [42].

The use of particles with multimodal particle size distribution and with small aspect ratios provided a relatively high maximum packing fraction, $\phi_m$ of 0.86. For all suspensions mixing index values were determined for each mixing condition to allow the documentation of the statistics of the spatial concentration variations of the ingredients of the suspension samples so that the suspension samples could be reproducibly prepared [42, 51, 52]. Mixing indices provided a quantitative understanding to enable the documentation of the effects of the processing conditions on the spatial homogeneity of ingredient distributions [42].

The suspensions with a $\phi$, range of 0.62 to 0.82 were characterized for their shear viscosity and wall slip behavior and it was shown that the rheological behavior can be represented by that of a generalized Newtonian fluid, with the shear viscosity material function represented well with the Herschel-Bulkley equation. The suspensions exhibited the apparent wall slip mechanism. Here, the parameters determined by He *et al.,* are employed, focusing on one of their suspensions, i.e., the suspension with a solids volume fraction, $\phi=0.76$ for the analysis of the steady torsional flow and the demonstration of the proposed method for the determination of the yield stress and other parameters [42]. The parameters of the Herschel-Bulkley equation and wall slip versus shear stress relationship are given in Table 1.

He *et al.*, used two different methods for the characterization of the yield stress value of the suspension at $\phi=0.76$ [42]. The first method involved the use of the straight marker line method, i.e., marking the free surface of the suspension and the edges of the two disks prior to the motion of the top disk [7, 41, 42] and then following the marker line as a function of time during steady torsional flow until steady state flow was achieved at each apparent shear rate. This method requires that the free surface of the sample be straight and that during the steady torsional flow the marker line on the free surface should not be broken (indicative of the fracture of the fluid during flow [7]). For viscoplastic fluids discontinuities in the marker line develop at both the top and bottom walls, indicating wall slip. For $|\tau_{z\theta}(r)| \leq \tau_0$ the suspension flows as a plug (as shown schematically in Fig. 1a) and for $|\tau_{z\theta}(r)| > \tau_0$ a deformation rate is imposed under steady state conditions (as shown schematically in Fig 1b). For all conditions apparent wall slip occurred at both walls. The second method that was used by He *et al.*, [42] for the determination of the yield stress values of their concentrated suspensions was based on the characterization of the wall slip velocity versus shear stress behavior of the suspensions and comparison of the wall slip velocities with the wall velocities [42]. As noted earlier, under plug flow in steady torsional flow the ratio of the slip velocity over the wall velocity becomes equal to 0.5. On the other hand, for $|\tau_{z\theta}(r)| > \tau_0$ the ratio of the slip velocity over the wall velocity becomes less than 0.5, with the transition enabling the determination of the yield stress values of the suspension samples [20, 42].

Both the straight marker line method and the wall slip analysis indicated that the yield stress, $\tau_0$, of the suspension consisting of a silicone polymer incorporated with multimodal particles at $\phi$=0.76 was equal to 85 Pa. The other parameters of the Herschel-Bulkley equation for the suspension, the shear viscosity of the binder and the thickness of the apparent slip layer, determined by He *et al.*, were the following (Table 1): Consistency index, $m$=2073 Pa-s$^n$, shear rate sensitivity index, n=0.8 (see equations 4c and 5a for definitions) [42]. The PDMS binder of the suspension was determined to exhibit Newtonian behavior $n_b$=1 for shear stresses < 3000 Pa with a viscosity of 4.9 Pa-s. The apparent slip layer thickness, $\delta$, at $\phi$=0.76 was determined to be a constant, i.e., $\delta$=0.6 μm (Table 1). Thus, a linear relationship should exist between the slip velocity and the shear stress for $|\tau_{z\theta}(r)| < 3000$ Pa, i.e., $U_s(r,0) = -U_s(r,H) = \frac{\delta}{m_b}(-\tau_{z\theta}(r))$.

**IV. Results and discussion:**

*Demonstration of the method for determination of the yield stress of viscoplastic fluids*

The methodology followed for the demonstration involved first the calculation of the torque values based on the Eq. (2a). To determine the torque values the parameters shown in Table 1 obtained by He *et al.*, were used for the calculation of the shear stress as a function of the radial direction, i.e., from $r=0$ to $r=R$, using Eq. (8a) and Eq. (8b) for various apparent shear rates at the edge, $\Omega R/H$ [42]. Eq. (8a) was solved incrementally in the radial direction for each set of $\Omega$ and $H$ via numerical approximation using the NSolve function of the Mathematica 11 kernel. The shear stress distributions, $\tau_{z\theta}(r)$, for various apparent shear rates at the edge at gap, $H=1$ mm, are shown in Fig. 2. As expected on the basis of the parameters of the shear viscosity of the viscoplastic suspension, the shear stress varies linearly with radial distance at locations where plug flow prevails and varies non-linearly in the deformation region.

Upon the calculation of the shear stress distribution, $\tau_{z\theta}(r)$, the torque, $\Im$, at each apparent shear rate was obtained via numerical integration, i.e., $\Im(H,\Omega) = 2\pi \int_0^R \left(-\tau_{z\theta}(r)\right) r^2 dr$. The calculated torque versus the apparent shear rate at the edge values for gaps, $H$, of 1 and 2 mm are shown in Fig. 3. For both gaps the slope $\dfrac{d\ln \Im}{d\ln(\Omega R/H)}$ changes at a critical torque, $\Im_c$ of $2.1*10^{-3}$ N-m. For both gaps up to this torque, i.e., $\Im < \Im_c$, the slope of the torque, $\Im(H,\Omega)$, versus the apparent shear rate at the edge, $\Omega R/H$, i.e., $\dfrac{d\ln \Im}{d\ln(\Omega R/H)} = 1$. On the other hand, the slope, $\dfrac{d\ln \Im}{d\ln(\Omega R/H)}$, becomes a function of the gap for $\Im > \Im_c$ i.e., $\dfrac{d\ln \Im}{d\ln(\Omega R/H)} = 0.7$ for $H=1$ mm, whereas the slope $\dfrac{d\ln \Im}{d\ln(\Omega R/H)} = 0.58$ for $H=2$ mm. Let us interpret the significant change in the slope $\dfrac{d\ln \Im}{d\ln(\Omega R/H)}$, at the critical torque, $\Im_c = 2.1*10^{-3}$ N-m (Fig. 3), i.e., going from a slope of 1 to 0.7 for $H=1$mm and from 1 to 0.579 for $H=2$mm.

What is the critical shear stress at the edge, $\left|\tau_{z\theta}(R)\right|_c$, that corresponds to the critical torque of $\Im_c =$ $2.1*10^{-3}$ N-m? Applying Eq. (3) for the critical condition, i.e.,

$$\left|\tau_{z\theta}(R)\right|_c = \frac{\mathfrak{I}_c}{2\pi R^3}\left(3 + \frac{d\ln\mathfrak{I}}{d\ln(\Omega R/H)}\right) \text{ with } \frac{d\ln\mathfrak{I}}{d\ln(\Omega R/H)} = 1.0 \text{ for } \mathfrak{I} < \mathfrak{I}_c \text{ the critical shear stress}$$

at the edge, $\left|\tau_{z\theta}(R)\right|_c$ is determined to be 85 Pa (Fig 3). As shown in Fig. 4 the slopes of the shear stress at the edge versus the apparent shear rate are similar for both gaps when the shear stress is less than 85 Pa. Above the critical shear stress the slopes depend on the gap. What occurs at the critical shear stress at the edge and what does 85 Pa represent?

As noted earlier, He *et al.* had determined (using two different methods) that the yield stress value of the suspension with ϕ=0.76 is 85 Pa [42]. Thus, the significant change in slope $\frac{d\ln\mathfrak{I}}{d\ln(\Omega R/H)}$ at the critical torque represents the transition from the plug flow regime, i.e., for $\left|\tau_{z\theta}(R)\right| \leq \tau_0$ to the continuous deformation regime, i.e., $\left|\tau_{z\theta}(R)\right| > \tau_0$ at which the suspension sustains a steady deformation rate at a given shear stress. As shown earlier in the plug flow region the slope $\frac{d\ln\mathfrak{I}}{d\ln(\Omega R/H)}$ is expected to be equal to 1 for a Newtonian binder (as expected from Eq. (9b)) and it is indeed 1 (Fig 3). For the continuous deformation region the analysis carried out for negligible slip in the continuous deformation rate region (Eq. (10d)) suggests that $\frac{d\ln\mathfrak{I}}{d\ln(\Omega R/H)} \approx n$. Considering that the shear rate sensitivity coefficient, *n*, of the suspension is 0.8 (see Table 1) the obtained values of 0.7 and 0.58 for gaps, H, of 1 and 2 mm reflect the effect of wall slip on the slope in the continuous deformation region. Overall, the transition in slope $\frac{d\ln\mathfrak{I}}{d\ln(\Omega R/H)}$ reflects a transition from plug flow regime to the deformation regime and presents the basis for the determination of the yield stress in steady torsional flow.

The determination of the yield stress requires that steady torque versus apparent shear rate data be collected under isothermal conditions over a sufficiently broad range of apparent shear rates to assure the inclusion of the transition between pure plug flow for $\left|\tau_{z\theta}(R)\right| < \tau_0$ (the flow within the gap is plug flow regardless of the radial position) and steady torsional flow involving a combination of plug flow plus continuous deformation, i.e., for $|\tau_{z\theta}(R)| > \tau_0$. The procedure below can be used so that a sufficiently broad range of apparent shear rates is applied.

- Collect the torque versus apparent shear rate data at multiple gaps. Inspect the torque versus apparent shear rate data. The torque versus apparent shear rate behavior should exhibit a change in slope at a critical torque value, $\Im_c$. This critical torque value should remain the same for the torque data collected at various gaps.

- This gap-independent transition in torque occurring at $\Im_c$ defines the yield condition for the viscoplastic fluid. A single gap is sufficient to define the yield condition and for the determination of the yield stress, $\tau_0$. However, multiple gaps are necessary for the characterization of the wall slip velocity versus the shear stress behavior.

- Obtain the derivatives $\dfrac{d\ln\Im}{d\ln(\Omega R/H)}$ for different gaps pertaining to torques that are smaller than the critical transition torque i.e., for $\Im < \Im_c$. When apparent wall slip mechanism prevails (and the apparent slip layer thickness, $\delta$, remains constant at different apparent shear rates) this derivative should remain constant for different gaps in the pure plug flow region, i.e., equal to $n_b$ for binders with a power law index, $n_b$ and equal to 1.0 for Newtonian binders, regardless of the gap, $H$, as long as $R >> H$. Convert all torque versus apparent shear rate data for the pure plug flow region collected with multiple gaps for which $\Im \leq \Im_c$ into $\tau_{z\theta}(R)$.

- Determine the derivatives $\dfrac{d\ln\Im}{d\ln(\Omega R/H)}$ for torques greater than the critical torque, i.e., $\Im > \Im_c$ for different gaps. Note that the value of this derivative for $\Im > \Im_c$ will be different for different gaps. Obtain the shear stress at the edge, $\tau_{z\theta}(R)$, values for $\Im > \Im_c$ using the derivatives, $\dfrac{d\ln\Im}{d\ln(\Omega R/H)}$ determined for different gaps.

- Using $\dfrac{d\ln \Im}{d\ln(\Omega R/H)}$ for the plug flow region and the critical torque, $\Im_c$ obtain the shear stress at the edge that corresponds to the critical torque, $\Im_c$. This shear stress is equal to the yield stress of the viscoplastic fluid, i.e., $|\tau_{z\theta}(R,\Im_c)| = \tau_0$.

As noted earlier, the steady torsional flow data can be collected at two [5] or more gaps [6] so that wall slip velocities at the edge can be determined as a function of shear stress. Over the plug flow region the deformation rates $\dfrac{dV_\theta}{dr}(\tau_{z\theta}(R))$ should be equal to zero and hence $U_s(R,0) = -U_s(R,H) = \Omega R/2$. The shear stress versus the true shear rate behavior and the slip velocity versus the shear stress behavior should be similar for different gaps.

Let us demonstrate these steps using the shear stress versus the apparent shear rate values shown in Fig. 4 employing the data collected at two gaps. Taking the derivative of both sides of Eq. (1b) with respect to reciprocal gap, $1/H$, gives the wall slip velocity at the shear stress at the edge (for two gaps, $H_1$ and $H_2$) [5]:

$$U_s(\tau_{z\theta}(R)) = \pm R \dfrac{\left[\dfrac{\Omega_1(\tau_{z\theta}(R))}{H_1} - \dfrac{\Omega_2(\tau_{z\theta}(R))}{H_2}\right]}{2\left(\dfrac{1}{H_1} - \dfrac{1}{H_2}\right)} \qquad (10)$$

where $\Omega_1(\tau_{z\theta}(R))$ and $\Omega_2(\tau_{z\theta}(R))$ are the rotational speeds for the two gaps at the same shear stress, $\tau_{z\theta}(R)$. As noted earlier, slip velocity at the stationary bottom wall (z=0) is positive, i.e., $U_s(R,0) = V_\theta(R,0)$, and the slip velocity at the moving top wall is negative, i.e., $U_s(R,H) = V_\theta(R,H) - \Omega R$. The true shear rate at the edge which corresponds to the shear stress at the edge is obtained from Eq. (1b) as [5]:

$$\dfrac{dV_\theta}{dz}(\tau_{z\theta}(R)) = R\dfrac{(\Omega_1(\tau_{z\theta}(R)) - \Omega_2(\tau_{z\theta}(R)))}{(H_1 - H_2)} \qquad (11)$$

Thus, knowing $\frac{dV_\theta}{dz}(\tau_{z\theta}(R))$ the shear viscosity of the fluid at the edge can be determined, i.e.,

$\eta(R) = |\tau_{z\theta}(R)| \Big/ \left[\frac{dV_\theta}{dz}(\tau_{z\theta}(R))\right]$. Although Yoshimura and Prud'homme have selected to work with two gaps, the analysis can be readily expanded to three of more gaps [6].

Once the shear stress at the edge, $|\tau_{z\theta}(R)|$, values are determined the slip velocity versus the shear stress relationship can be determined using Eq. (10). The application of Eq. (10) to the shear stress at the edge versus the apparent shear rate data (Fig. 4) generated the wall slip velocity results shown in Fig. 5, which accurately reproduce the apparent slip behavior of the suspension (the slip coefficient, i.e., $\beta = \frac{\delta}{\mu_b} = \frac{6 \cdot 10^{-7}}{4.9} = 1.2 \cdot 10^{-7}$).

To validate the yield stress value obtained with the torque versus the apparent shear rate data one can also probe the relationship between the wall slip velocity and the velocity of the disk driving the steady torsional flow. The ratio of the wall slip velocity over the velocity of the top plate at the edge, i.e., $\frac{|U_s|}{\Omega R}$ versus the shear stress at the edge, $(-\tau_{z\theta}(r))$, is shown in Fig. 6. Plug flow is indicated when the ratio $\frac{|U_s|}{\Omega R} = 0.5$ [20, 42]. The transition between the plug flow and the continuous deformation flow region is the yield stress of the suspension, which is indicated to be 85 Pa. Thus, the yield stress value determined from wall slip analysis agrees with the yield stress determined from torque versus the apparent shear rate data. The effect of the surface to volume ratio of the rheometer is shown in Fig. 6. As expected, the effect of wall slip is more significant at the smaller gap, i.e., greater surface to volume ratio.

The true shear rates at the edge, $\frac{dV_\theta}{dz}(\tau_{z\theta}(R))$, can be determined using Eq. (11). The shear stress at the edge versus the true shear rate at the edge for the two gaps are shown in Fig. 7. Since the yield stress value of the suspension is determined the other two parameters of the Herschel-Bulkley equation can be determined using the shear stress at the edge versus the true (slip corrected) shear rate at the edge curves. The best fit with m=2073 Pa-s$^n$ and n=0.8 is shown in Fig. 7 and reproduces accurately the flow curve of the suspension. Overall, it is thus demonstrated that the yield stress value of a viscoplastic fluid can be determined using only the torque versus the apparent shear rate data, and these data can be generated relatively easily employing the steady torsional flow, with the rest of the parameters of shear viscosity and wall slip obtained with conventional methods.

There are a number of approximating procedures, "single point" methods, that aim to extract estimates of the shear viscosity of non-Newtonian fluids using the apparent shear viscosity data [52-54]. The basic premise is that the true and apparent shear stress become equal to each other at a "representative position" which for steady torsional flow is observed to be at $r$=0.76$R$ [54]. It is indicated that at the representative position the shear viscosity would approximate the apparent shear viscosity within ±2%. In the Appendix these methods are described and typical comparisons of the distributions of the true shear rate, shear stress and shear viscosity and their comparisons with their apparent values are provided for a range of apparent shear rate at the edge values (Fig. 8-10). Although the apparent and true shear stress values indeed become equal around $r$=0.76$R$ the apparent shear rates are significantly greater than the true shear rates and consequently the true shear viscosity values are higher, with orders of magnitude differences at shear stresses which approach the yield stress.

There are interesting ramifications of these steady torsional flow results on the cone and plate flow. It is demonstrated here that regardless of the radial position the flow in between the two disks can be treated as occurring between infinitely wide and long parallel plates for a viscoplastic suspension with apparent slip. This suggests that a similar approach for viscoplastic fluids can be applicable to the cone and plate flow. The adequacy of the parallel plate treatment for cone and plate flow has indeed been demonstrated on the basis of confocal microscopy for viscoplastic refractive index matched colloidal dispersions of rigid poly(methyl methacrylate) nanoparticles (radii of 138-302 nm) incorporated into a mixture of two Newtonian fluids [55]. Future investigations of the cone and plate flow can thus use the parallel plate treatment to determine the constraints related to its use for the characterization of complex fluids including viscoplastic fluids subject to wall slip.

## V. Conclusions

The steady torsional flow occurring in between two parallel disks was analyzed using the earlier-determined parameters of the shear viscosity and wall slip of a viscoplastic suspension of multi-modal rigid particles incorporated into a Newtonian fluid [42]. It is demonstrated that the steady torsional flow, occurring under creeping and isothermal flow conditions in between two disks (radius, $R\gg$ gap, $H$), can be treated as flow occurring in between series of infinitely-wide and infinitely-long parallel plates with the wall velocity values of $\Omega r$. The torque, $\Im$, data obtained from the integration of the shear stress distributions at each rotational speed, $\Omega$, indicated that there is not one but minimum of two slopes, $\dfrac{d\ln\Im}{d\ln(\Omega R/H)}$ below and above a critical torque. The critical torque is independent of the gap of the rheometer. Plug flow prevails below the critical torque at which the slope $\dfrac{d\ln\Im}{d\ln(\Omega R/H)}$ is independent of the gap. Above the critical torque the slope $\dfrac{d\ln\Im}{d\ln(\Omega R/H)}$ is a function of the gap. The slopes can be used to convert the torque data to true shear stress at the edge of the disk for both the plug flow which occurs below the critical torque and for the deformation regime above the critical torque. The shear stress which corresponds to the critical torque is the yield stress of the viscoplastic fluid. The ability to thus determine the yield stress using only a single gap offers significant advantages. The determination of the yield stress further allows the unambiguous characterization of the other parameters of the shear viscosity material function as well as the wall slip velocity versus shear stress behavior of the viscoplastic fluid using multiple gaps. The gap-dependent data can be used to verify the yield stress value of the viscoplastic fluid via the comparisons of the wall slip velocity values with the wall velocities.

The parallel plate analysis of steady torsional flow allows the determination of the radial distributions of the true values of the shear stress, shear rate and shear viscosity. These distributions and their comparisons with apparent data suggest that one-point methods relying on the matching of the apparent and true shear stress values at a representative location of $r=0.76R$ [52-54] are not applicable for viscoplastic fluids subject to wall slip when the shear stress at the edge values are not far removed from the yield stress.

Overall, these results further reinforce the accuracy and practicality of a paradigm shift away from the use of roughened rheometer surfaces, that frequently leads to the fracture of viscoplastic samples during viscometric flows, to the use of smooth surfaces for the concomitant characterizations of the wall slip behavior and the shear viscosity of viscoplastic fluids. The procedures outlined here for steady torsional flow can be readily integrated to the analysis software of rotational rheometers for an accurate analysis and the determination of the parameters of wall slip behavior and the shear viscosity material function of viscoplastic fluids.

## VI. Acknowledgments

This paper is dedicated to the memory of Dr. Joseph Starita, the founder of Rheometric Scientific (currently TA Instruments), whose friendship I very much valued since the time when our paths crossed at Stevens starting in 1980s.

I thank Prof. Morton Denn for suggesting the analysis of the steady torsional flow and Ms. Li Quan for her help in the preparation of the figures.

## VII. Appendix
**Comparison of shear viscosity results with those from one step methods**

There are various methods that were proposed for the determination of the shear viscosity material function of various non-Newtonian fluids from their apparent shear viscosity data. The major approaches are generally named as "single point methods" [ 52-54].

A simple case can be used to demonstrate how these methods work. For example, for steady, fully-developed, isothermal and laminar flow of power law fluids in a capillary die the shear stress distribution is linear with radial position, r, both for Newtonian and power-law fluids. From analytical solutions it is known that the apparent shear rate and true shear rate curves cross each other at a radial position that is a function of the power-law index n. A good estimate of the true shear viscosity can be obtained at a representative shear rate that is equal to 83% of the apparent shear rate at the wall, at which location the true shear viscosity becomes equal to the apparent shear viscosity within ±2%) [52, 54].

Similarly, for the steady torsional flow it is asserted that the true shear viscosity of a non-Newtonian fluid would be equal to the apparent shear viscosity of the fluid at a radial position at which the shear stress is equal to 76% of the apparent shear stress [54], hence at a representative position of r=0.76R. A complete analysis of the one-point method on the basis of the apparent shear stress versus the true shear stress, the apparent shear rate versus the true shear rate and hence the apparent shear viscosity versus the true shear viscosity was made and typical results for $H$=1 mm are shown in Fig. 8-10. The true shear stress values are equal to each other in the plug flow region. In the continuous deformation region the differences between the true and apparent shear stress values are also not too far removed from each other as shown in Fig. 8. The true shear stress values are greater than the apparent shear stress values for small r, but approach each other and become equal around r=0.8R. This is reasonable considering that the use of $\frac{d\ln\Im}{d\ln(\Omega R/H)}=1$ versus the true values of the slopes as 0.70 or 0.58 (Fig. 3) introduce only changes of about 10% in the determined torque value. Similarly, the determination of and use of a single slope $\frac{d\ln\Im}{d\ln(\Omega R/H)}$ to cover both the plug flow regime and the continuous deformation region or using two separate slopes would make relatively small differences on the shear stress at the edge values that are obtained. If all of the torque versus apparent shear rate data of Fig. 3 were to be considered altogether to generate a single slope, $\frac{d\ln\Im}{d\ln(\Omega R/H)}=0.906$ for $H$=1mm and $\frac{d\ln\Im}{d\ln(\Omega R/H)}=0.831$ for $H$=2mm. Using a single slope versus two slopes would introduce errors only in the order of -2 to 6% for $H$=1mm and -4 to 7% (low to high apparent shear rate) in the determination of the true shear stress at the edge values.

However, the values of the true shear rate are very different than those of apparent shear rate, as shown in Fig. 9. For the plug flow zones the true shear rate is zero, whereas the apparent shear rates at the same locations are finite. For the continuous deformation regions again the true shear rate values are significantly smaller than the apparent shear rate values. Consequently, the true and apparent shear viscosity values are very different (Fig. 10). The true shear viscosity values are infinite, indicating solid-like behavior characteristic of the plug flow region when the shear stress is smaller than the yield stress. The true shear viscosities are significantly greater than the apparent shear viscosity values in the deformation region for relatively small r/R. As the edge is approached the true and apparent shear viscosity values approach each other. However, they never cross. Thus, caution must be exercised when the one step methods are to be used for viscoplastic fluids as a result of the decreases in shear rate associated with their ubiquitous wall slip behavior.

**List of figures:**

1. Schematics of the steady torsional flow (a) plug flow with apparent slip (b) deformation with apparent slip at the wall

2. Distributions of the shear stress versus r/R at the apparent shear rate at the edge, $\Omega R/H$, range of 0.005 to 0.2 s$^{-1}$.

3. Steady torque, $\Im$, versus the apparent shear rate at the edge, $\Omega R/H$, for two gaps. The critical torque corresponds to the yield condition from which the yield stress can be determined. The calculations for plug flow condition (unbroken lines) were carried out using Eq. 9.

4. Shear stress at the edge, $|\tau_{z\theta}(R)|$, versus the apparent shear rate at the edge, $\dot{\gamma}_{aR} = \Omega R/H$, for two gaps, H=0.1 and 0.002 m.

5. Wall slip velocity, |Us|, versus shear stress at the edge, $|\tau_{z\theta}(R)|$, from the data collected at the two gaps. The slope is equal to 1.22*10$^{-7}$, m/(Pa-s).

6. Wall slip velocity, |Us|, over the plate velocity ($\Omega$*R) versus shear stress at the edge, $|\tau_{z\theta}(R)|$, from the data collected at the two gaps. |Us|/($\Omega$R)=0.5 corresponds to plug flow.

7. Shear stress at the edge, $|\tau_{z\theta}(R)|$, versus the true (slip corrected) shear rate at the edge, $\frac{dV_\theta}{dr}(R) = \dot{\gamma}(R)$, from the data collected at the two gaps, H. The parameters of Herschel-Bulkley equation are yield stress, $\tau_0$=85 Pa, m=2073 Pa-s$^n$ and n=0.8.

8. Radial distributions of apparent shear stress, $|\tau_{z\theta,app}(r)|$ (unfilled) and true shear stress $|\tau_{z\theta}(r)|$ (filled symbols) for the apparent shear rates at edge, $\dot{\gamma}_{aR} = \Omega R/H$, of 0.05, 0.1 and 0.2 s$^{-1}$ (H=1mm).

9. Radial distributions of true shear rate, $\frac{dV_\theta}{dr}(r) = \dot{\gamma}(r)$, (filled symbols) and apparent shear rate $\dot{\gamma}_{ar} = \Omega r/H$ (unfilled) in the apparent shear rate at the edge range 0.02 to 0.2 s$^{-1}$ (H=1mm).

10. Fig. 10 Apparent shear viscosity, $\eta_{apparent}$ (unfilled symbols) and true shear viscosity, $\eta_{true}$ (filled symbols) distributions as a function of r/R at the apparent shear rates of 0.05, 0.1 and 0.2 $s^{-1}$ (H=1mm).

**Table 1:** Parameters of wall slip velocity versus shear stress and shear viscosity as determined by He *et al.,* 2019. These parameters were used to generate the torque versus the apparent shear rate behavior shown in Fig. 2.

Shear viscosity: (Herschel-Bulkley see Equations 4 and 5)
Consistency index, $m = 2073$ Pa-s$^n$,
Shear rate sensitivity index, n=0.8
Yield stress, $\tau_0 = 85$ Pa

Apparent wall slip behavior (see Equation 7 for definitions)
For shear stresses < 3000 Pa: $n_b = 1$ and $m_b = \mu_b = 4.9$ Pa-s.
Slip layer thickness, $\delta = 0.6$ μm

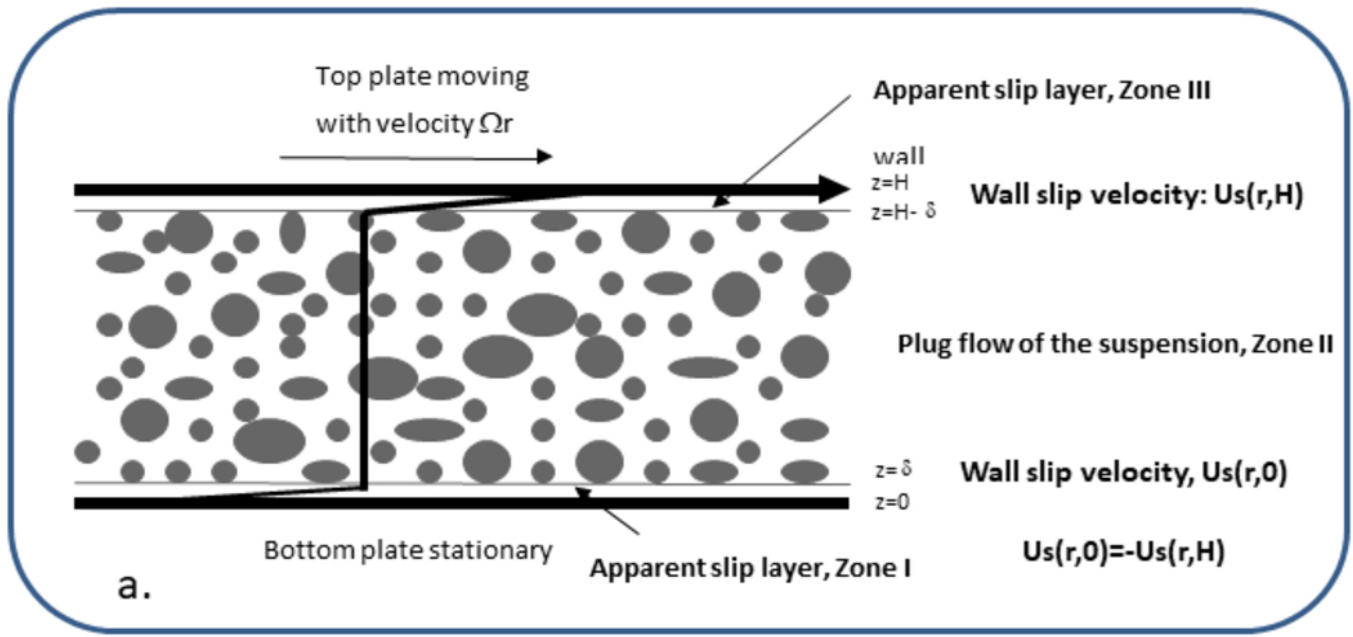

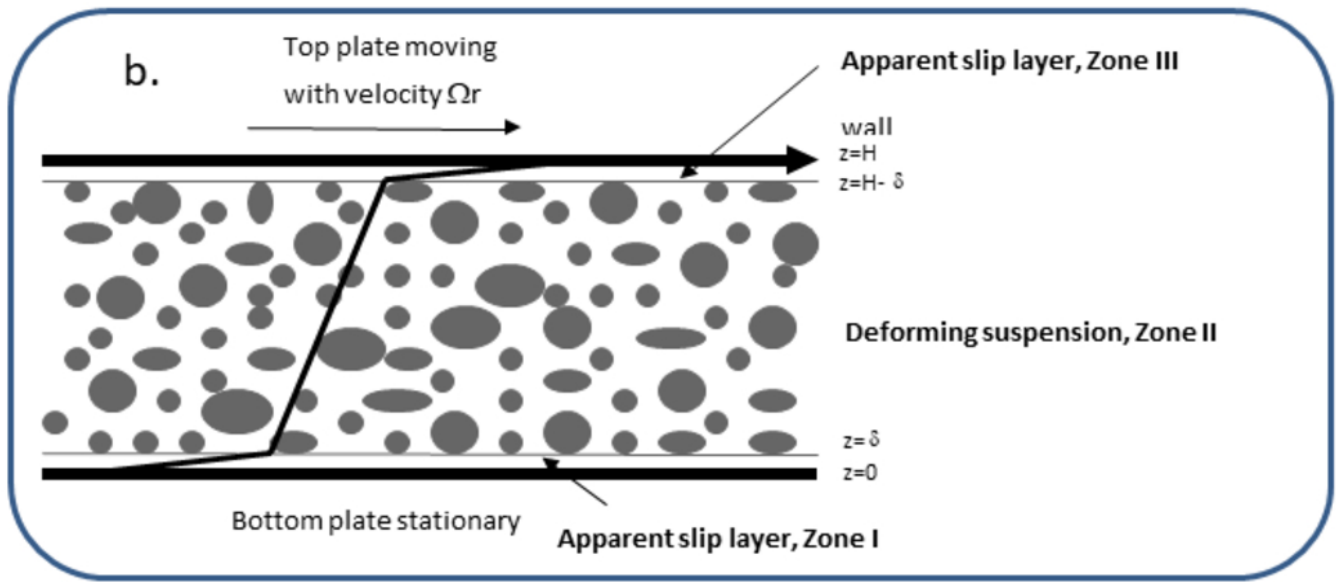

Figure 1: Schematics of the steady torsional flow (a) plug flow with apparent slip (b) deformation with apparent slip at the wall

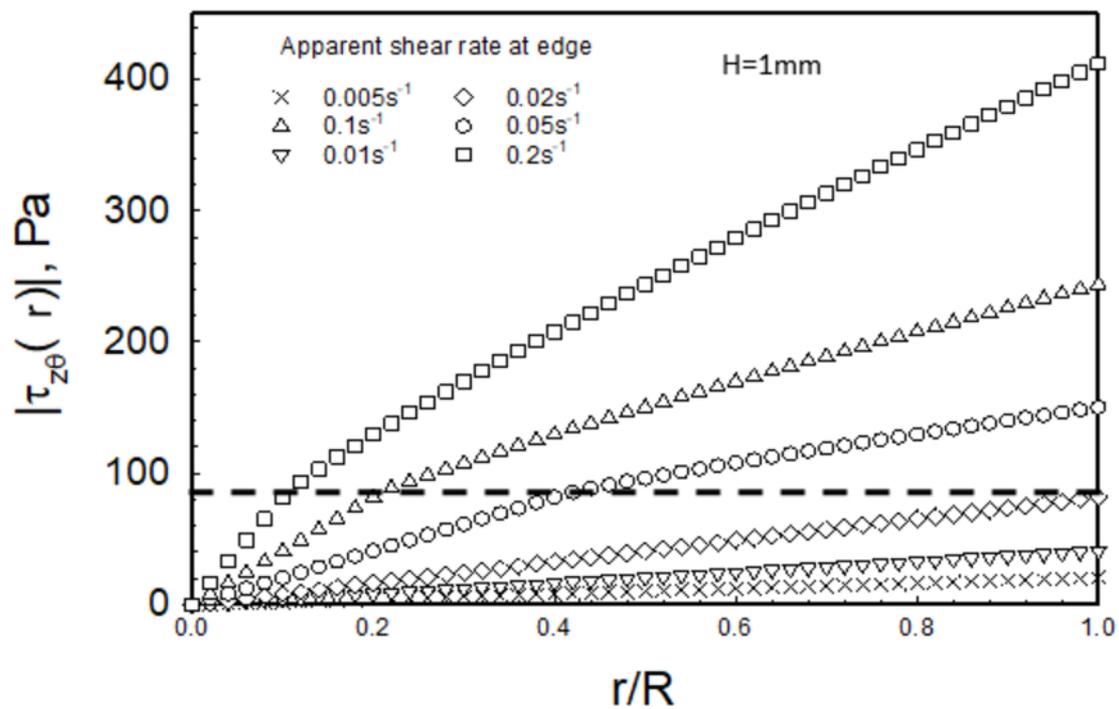

Figure 2: Distributions of the shear stress versus r/R at the apparent shear rate at the edge range of 0.005 to 0.2 s$^{-1}$.

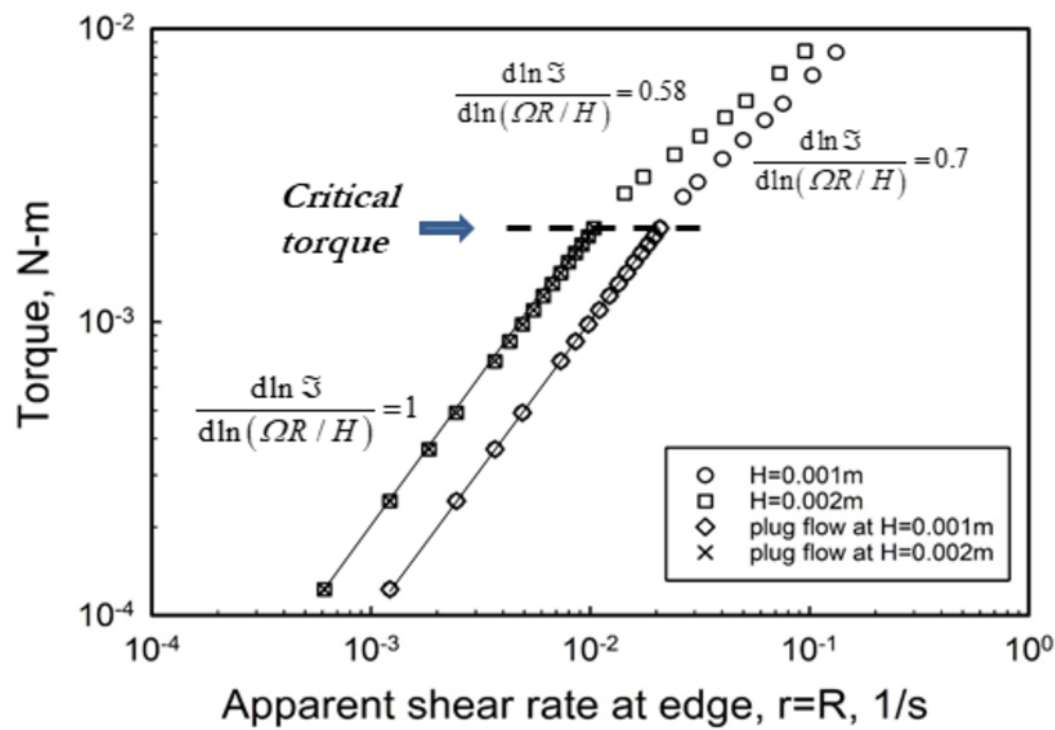

Figure 3. Steady torque, $\mathfrak{I}$, versus the apparent shear rate at the edge, $\Omega R/H$, for two gaps. The critical torque corresponds to the yield condition from which the yield stress can be determined. The calculations for plug flow condition (unbroken lines) were carried out using Eq. 9.

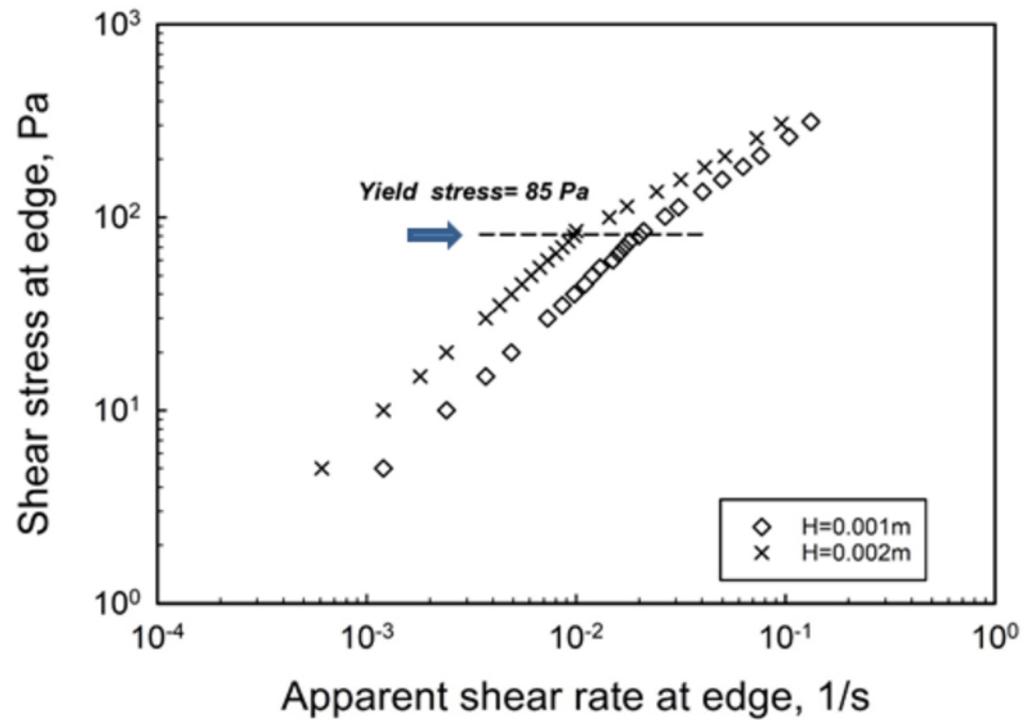

Figure 4: Shear stress at the edge, $|\tau_{z\theta}(R)|$, versus the apparent shear rate at the edge, $\dot{\gamma}_{aR} = \Omega R / H$, for two gaps, H=0.001 and 0.002m.

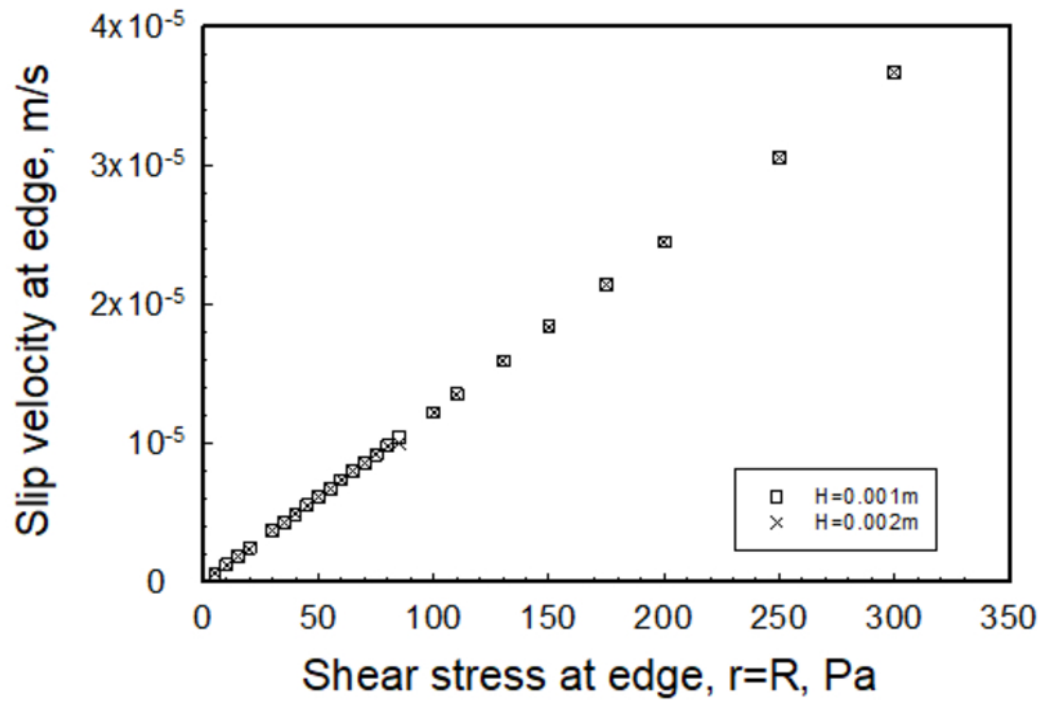

Figure 5: Wall slip velocity, $|U_s|$, versus shear stress at the edge, $|\tau_{z\theta}(R)|$, from the data collected at the two gaps. The slope is equal to $1.22*10^{-7}$ m/(Pa-s).

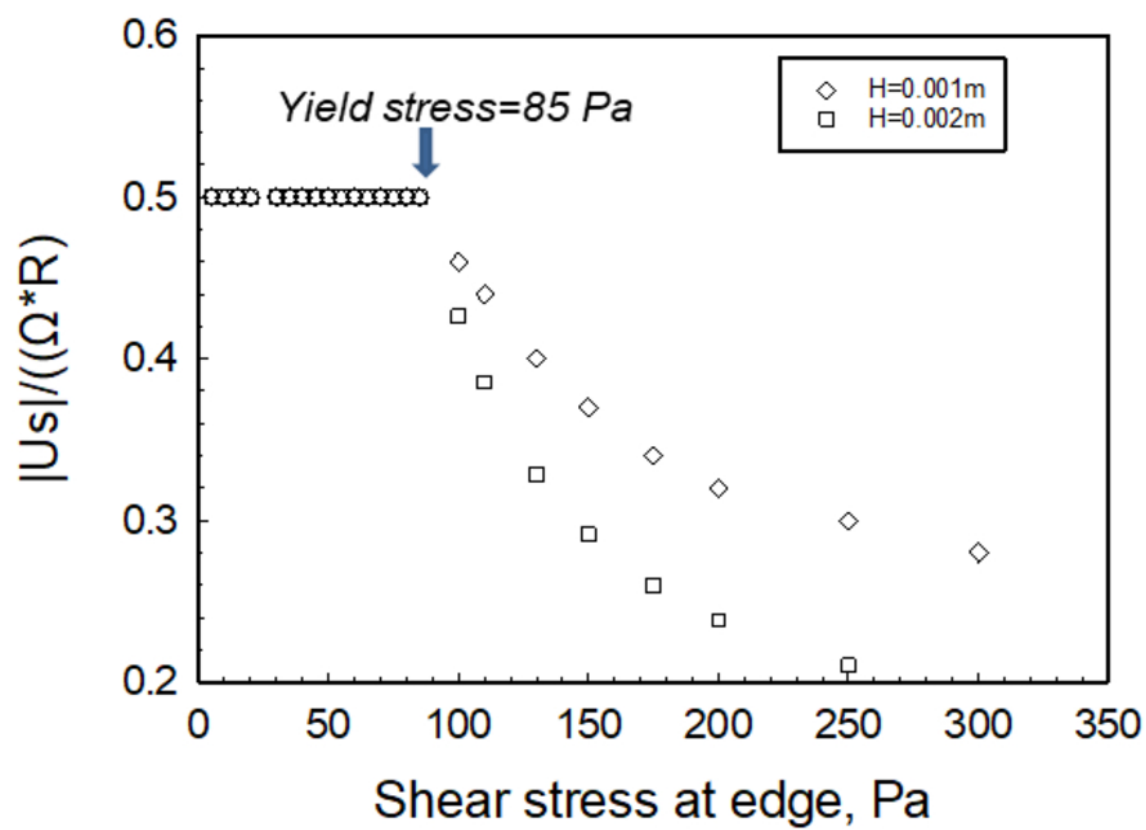

Figure 6: Wall slip velocity, |Us|, over the plate velocity (Ω*R) versus shear stress at the edge, $|\tau_{z\theta}(R)|$, from the data collected at the two gaps. |Us|/(ΩR)=0.5 corresponds to plug flow.

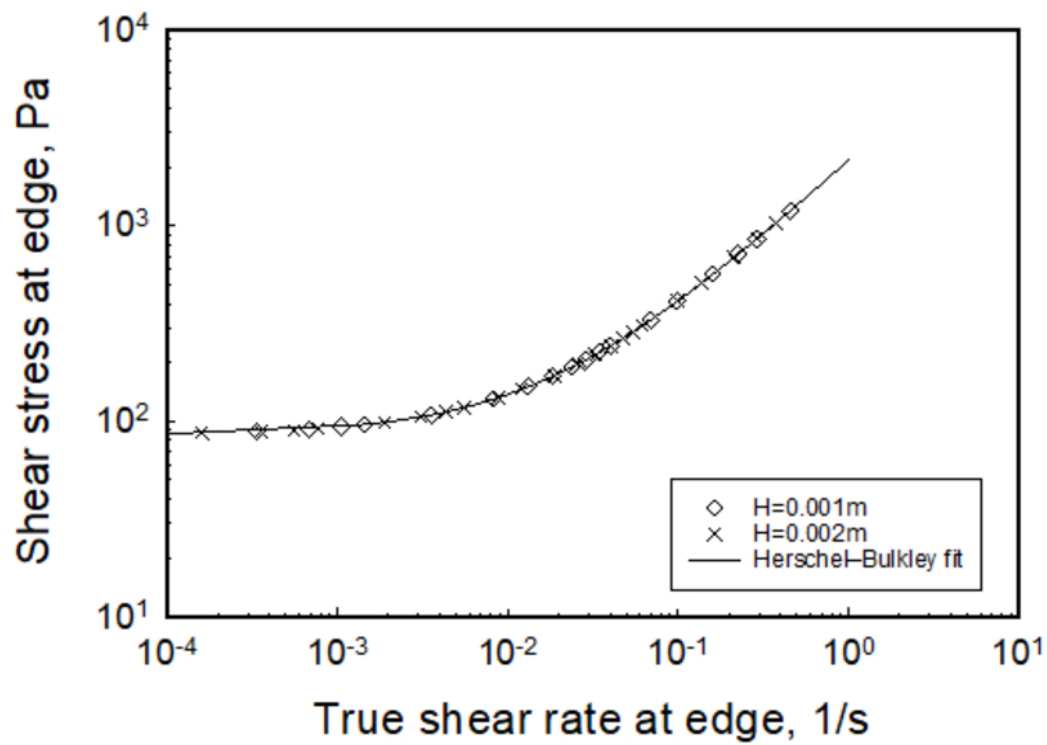

Figure 7: Shear stress at the edge, $|\tau_{z\theta}(R)|$, versus the true (slip corrected) shear rate, $\frac{dV_\theta}{dr}(R)$, at the edge From the data collected at the two gaps. The parameters of Herschel-Bulkley equation are: yield stress, $\tau_0$ =85 Pa, m=2073 Pa-s$^n$ and n=0.8.

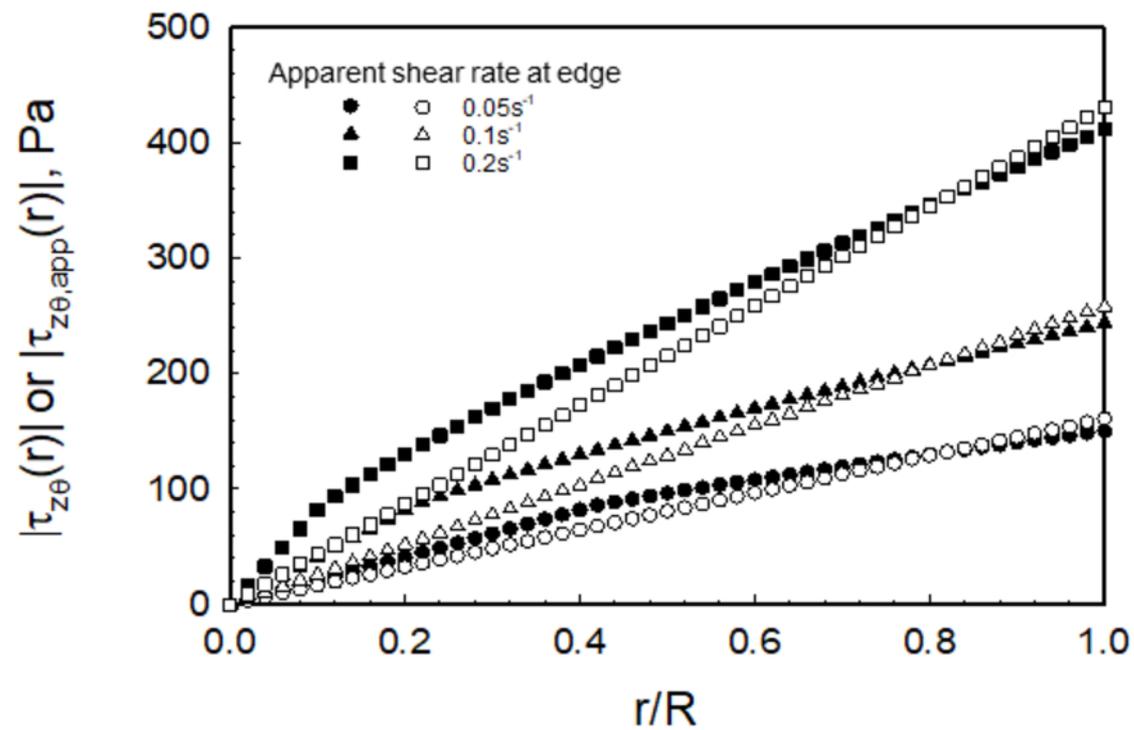

Fig. 8 Radial distributions of apparent shear stress (unfilled) and true shear stress (filled symbols) for the apparent shear rates at edge of 0.05, 0.1 and 0.2 s$^{-1}$ (H=1mm).

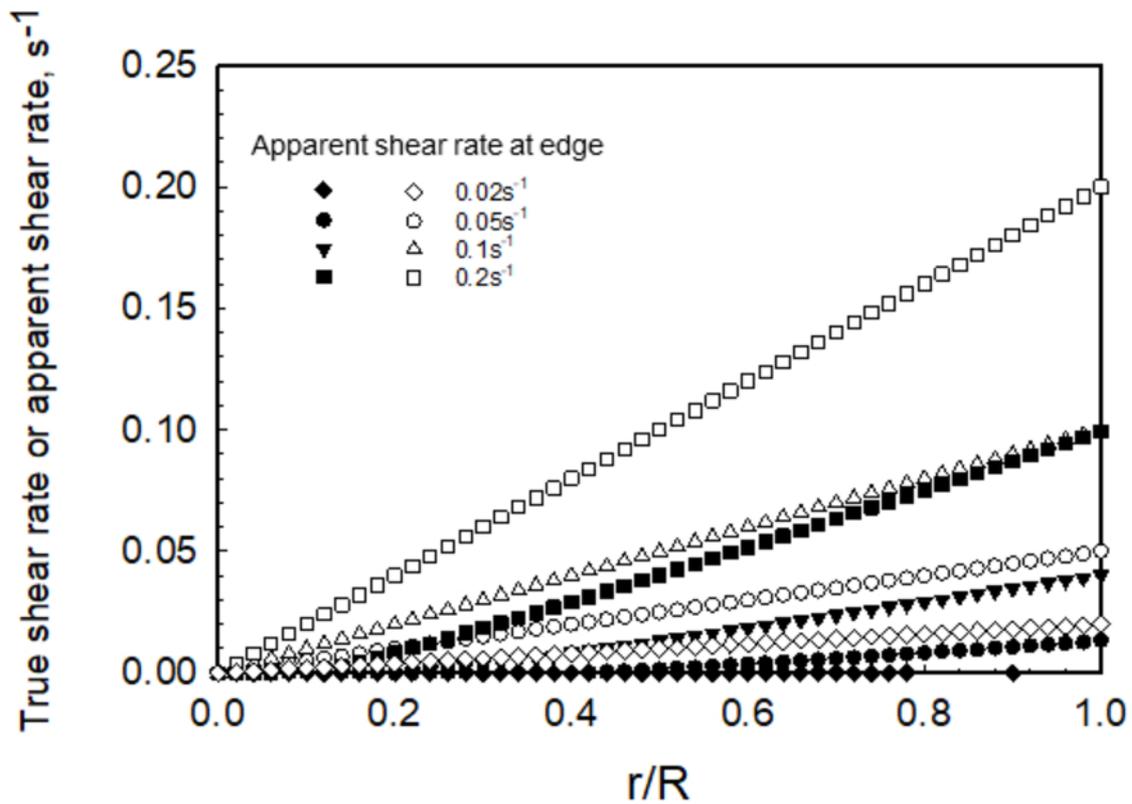

Fig. 9 Radial distributions of true shear rate (filled symbols) and apparent shear rate (unfilled) in the apparent shear rate at the edge range 0.02 to 0.2 s$^{-1}$ (H=1mm).

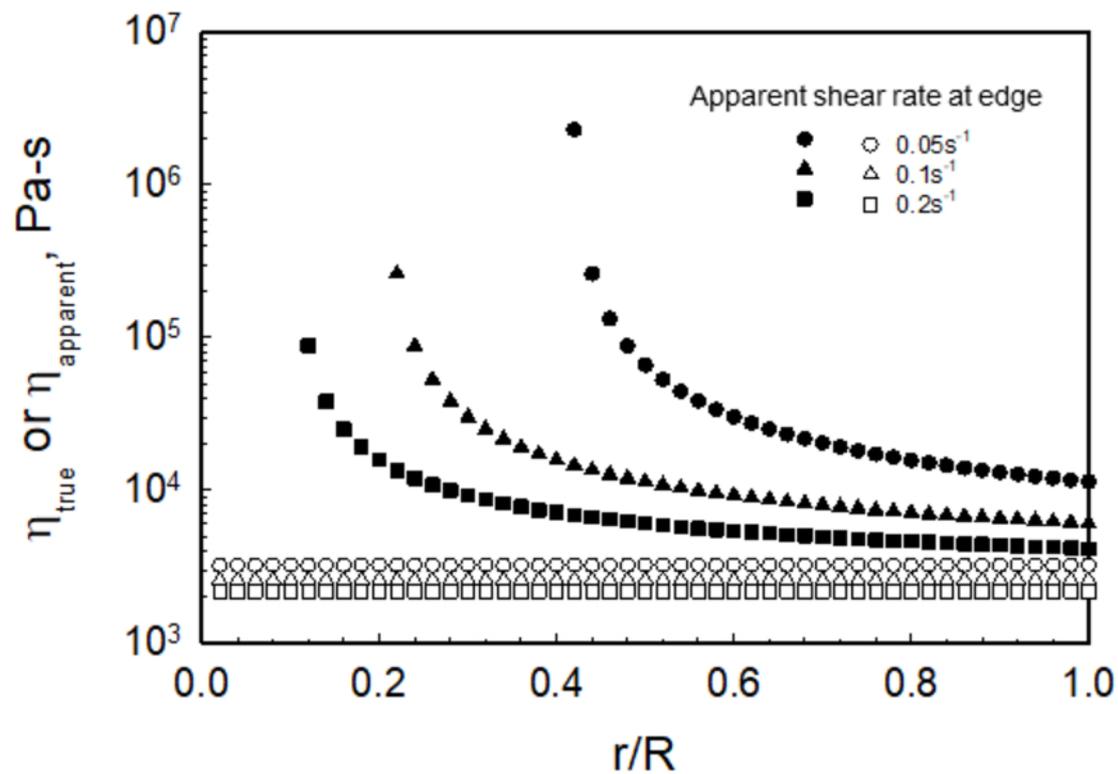

Fig. 10 Apparent shear viscosity, $\eta_{apparent}$ (unfilled symbols) and true shear viscosity $\eta_{true}$ (filled symbols) distributions as a function of r/R at the apparent shear rates of 0.05, 0.1 and 0.2 s$^{-1}$ (H=1mm).